\documentclass[
]{ceurart}

\usepackage[rgb]{xcolor}

\selectcolormodel{natural}
\usepackage{ninecolors}
\selectcolormodel{rgb}
\usepackage{tabularray}

\usepackage{listings}
\usepackage{booktabs}
\usepackage{tabularx}
\usepackage{caption}
\usepackage{subcaption}
\begin{document}

\copyrightyear{2022}
\copyrightclause{Copyright for this paper by its authors.
  Use permitted under Creative Commons License Attribution 4.0
  International (CC BY 4.0).}

\conference{LLM4Eval: The First Workshop on Large Language Models for Evaluation in Information Retrieval, 18 July 2024, Washington DC, United States}

\title{Using LLMs to Investigate Correlations of Conversational Follow-up Queries with User Satisfaction}

\tnotemark[1]
\tnotetext[1]{You can use this document as the template for preparing your
  publication. We recommend using the latest version of the ceurart style.}

\author[1]{Hyunwoo Kim}[
orcid=0000-0002-4923-6650,
email=khw0726@kaist.ac.kr
]
\fnmark[1]
\author[1]{Yoonseo Choi}[
orcid=0000-0001-6808-2848,
email=yoonseo.choi@kaist.ac.kr
]
\fnmark[1]
\author[2]{Taehyun Yang}[
email=0705danny@snu.ac.kr
]
\author[3]{Honggu Lee}[
email=honggu.lee@navercorp.com
]
\author[3]{Chaneon Park}[
email=chaneon.park@navercorp.com
]
\author[3]{Yongju Lee}[
email=lee.yongju@navercorp.com
]
\author[3]{Jin Young Kim}[
email=jin.y.kim@navercorp.com
]
\author[1]{Juho Kim}[
orcid=0000-0001-6348-4127,
email=juhokim@kaist.ac.kr
]
\cormark[1]
\address[1]{School of Computing, KAIST, 291 Daehak-ro, Yuseong-gu, Daejeon, 34141, Republic of Korea}
\address[2]{College of Liberal Studies, Seoul National University, 1 Gwanak-ro, Gwanak-gu, Seoul, 08826, Republic of Korea}
\address[3]{Naver Corp., 95 Jeongjail-ro, Bundang-gu, Seongnam-si, Gyeonggi-do, 13561, Republic of Korea}

\cortext[1]{Corresponding author.}
\fntext[1]{These authors contributed equally.}


\begin{abstract}
With large language models (LLMs), conversational search engines shift how users retrieve information from the web by enabling natural conversations to express their search intents over multiple turns. 
Users' natural conversation embodies rich but implicit signals of users' search intents and evaluation of search results to understand user experience with the system.  
However, it is underexplored how and why users ask follow-up queries to continue conversations with conversational search engines and how the follow-up queries signal users' satisfaction. 
From qualitative analysis of 250 conversational turns from an in-lab user evaluation of Naver Cue:, a commercial conversational search engine, we propose a taxonomy of 18 users' follow-up query patterns from conversational search, comprising two major axes: (1) users' motivations behind continuing conversations (N = 7) and (2) actions of follow-up queries (N = 11). 
Compared to the existing literature on query reformulations, we uncovered a new set of motivations and actions behind follow-up queries, including asking for subjective opinions or providing natural language feedback on the engine's responses. 
To analyze conversational search logs with our taxonomy in a scalable and efficient manner, we built an LLM-powered classifier (73\% accuracy).
With our classifier, we analyzed 2,061 conversational tuples collected from real-world usage logs of Cue: and examined how the conversation patterns from our taxonomy correlates with satisfaction.  
Our initial findings suggest some signals of dissatisfactions, such as \textbf{Clarifying Queries}, \textbf{Excluding Condition}, and \textbf{Substituting Condition} with follow-up queries. 
We envision our approach could contribute to automated evaluation of conversation search experience by providing satisfaction signals and grounds for realistic user simulations.
\end{abstract}

\begin{keywords}
  conversational search \sep
  follow-up queries \sep
  search intent \sep
  search evaluation
\end{keywords}

\maketitle

\section{Introduction}

The evolution of large language models (LLMs) is reshaping how people search for information, which was traditionally centered around information retrieval using users' keyword-based queries. 
% Conversational search engines provide selectively summarized information as search results to provide a natural conversational experience~\cite{liao2020Conversational}, encouraging users' multi-turn queries to explore broader and deeper information. 
% % With LLMs' capabilities for natural conversations, conversational search engines allow users to express and refine their search intents through multiple conversational queries~\cite{liao2020Conversational}.
With LLM's capabilities to understand natural language queries, users can retrieve information in a more natural manner. Furthermore, users can react to the responses of the search engine with natural language utterances to express and refine their search intents, explore the search results, and give feedback to the search results~\cite{liao2020Conversational}.
Compared to the fixed behavioral signals (e.g., clicks, reformulations, dwell time) with conventional keyword-based search, users' natural language conversation provides richer signals about users' search intents and evaluation of search results.
However, the complexity of natural language utterances makes it challenging for search designers and developers to discern the users' search intent and evaluate responses from the conversation. 
% (윤서) get the right intent of users since the natural language express more things in the search queries but still imply hidden user intent. 
Furthermore, the user can evolve their search intent from interacting with the search engine, posing additional challenges to understanding the dynamic user intent within the conversation session. 
% as users' experience with search engines gets more complex spanning multiple turns, the user intent during the entire conversation session can evolve, which makes it challenging to understand user experience and evaluate users' satisfaction for both single turn of query-response and holistic conversation session. 

% However, the introduction of conversational search brings significant challenges in understanding user experience and evaluating users' satisfaction due to the increased complexity of user experience across multiple turns of search.

% Evaluating interactions with conversational search needs to be holistically reviewed from two aspects: search and conversational interaction. 
There has been a large body of work on understanding and evaluating user experience with user interactions from the perspectives of information retrieval (IR) and conversational interactions. 
Previous IR studies have documented users' interaction patterns signaling users' satisfaction, such as clicks on search results~\cite{chapelle2012Largescale}, query reformulations~\cite{hassan2013clicks, jansen2007Patterns, huang2009Analyzing}, and query abandonments~\cite{diriye2012Leaving}.
Such behavioral patterns can complement offline evaluations with insights into actual users' evaluation of search results at scale~\cite{hofmann2016Online}. However, it's still underexplored how behavioral patterns should be defined and analyzed within the context of conversational searches. 
% Previous informational retrieval (IR) studies have documented interaction patterns in traditional search contexts~\cite{diriye2012Leaving, hassan2013clicks}. these patterns do not directly translate into conversational search context. 
% Unlike traditional search engines, where satisfaction or dissatisfaction may inferred from simplified interaction metrics such as click-through rates and query reformulation patterns~\cite{chen2021Better, huang2009Analyzing, jansen2007Patterns} and abandonments~\cite{diriye2012Leaving}, conversational search systems demand further understanding of multi-turn dialogue progression and user intent behind the queries. 
In terms of conversational interaction, existing work has explored diverse interaction scenarios with conversational agents to understand interaction patterns of satisfaction/dissatisfaction~\cite{myers2018Patterns, see2021Understanding, kim2023Understanding} and predict user satisfaction~\cite{choi2019Offline, park2020Largescale, jiepu2015automatic}.
% However, as they focused on generic, multi-purpose conversational interactions, these insights have not been thoroughly connected with user satisfaction with search tasks. 
At the intersection of information search and conversational interactions, a thread of research investigated and conceptualized interaction patterns for information-seeking conversations~\cite{Vakulenko2019QRFA, qu2018analyzing, Azzopardi2018, radlinski2017Theoretical}.
However, while existing literature on the evaluation of search systems and conversational interactions points out that interaction patterns with conversational search engines can be rich and meaningful signals for the evaluation of conversational search systems, the relationship between the interaction patterns and user satisfaction within the domain of conversational information retrieval is still underexplored. 
% Unlike traditional search engines, where satisfaction or dissatisfaction may inferred from simplified interaction metrics such as click-through rates and query reformulation patterns~\cite{chen2021Better, huang2009Analyzing, jansen2007Patterns} and abandonments~\cite{diriye2012Leaving}, conversational search systems demand further understanding of multi-turn dialogue progression and user intent behind the queries. 

In this research, we aim to answer two questions:
\begin{itemize} 
    \item Q1. How do the user interaction patterns in conversational search differ from those in conventional search?
    \item Q2. How do the user interaction patterns correlate with the user satisfaction of the conversational search? 
\end{itemize}

% To initially focus on understanding the distinct user interaction patterns shown in conversational search usages
To answer Q1, we analyzed how the users interact with conversational search queries. 
With nine predefined queries, we collected a total of 96 sessions of conversational search from 72 lay users by letting them freely complete the search tasks with Naver Cue:~\footnote{https://cue.search.naver.com/}. 
To compare the interaction patterns with those from conventional searches in literature, we focused on `follow-up queries' for analysis, which reveals how the users actively steer their search tasks. 
We qualitatively analyzed 250 conversational tuples (the set of `User Query N - System Response to N - User Query N+1') from collected conversations and organized the patterns of follow-up queries in the form of taxonomy. 
In total, we organized 18 themes across two axes: (1) users' motivations behind continuing the conversations (N = 7) and (2) patterns of follow-up queries (N = 11).

% With taxonomy, we aimed to automate the classification process of conversation to enhance scalability and expandability. 
% To analyze conversation logs at scale, 
To answer Q2, we developed an automated process to classify conversations using our taxonomy.
We first developed an LLM-powered classifier with our proposed taxonomy to analyze conversations at scale. 
We tested the classifier with ground truth labels from the taxonomy creation and observed 73\% accuracy. 
With the classifier, we analyzed 2,061 conversational tuples sampled from real-world usage logs of Naver Cue:, with ground-truth satisfaction ratings from external evaluators.
% a commercial service of a conversational web search engine.
% , to evaluate the user experience from conversation search logs at scale. %윤서: classifier 이야기를 더 하고싶음 
% to demonstrate the generalizability and to examine the correlation between our taxonomy and search satisfaction.
% The results showed that the classifier could successfully classify more than 80\% of the conversational tuples. 
From comparing the patterns of follow-up queries between different levels of search satisfaction, we observed that unsatisfactory conversations involve clarifying queries and reacting to search engine's responses more often.
We envision our taxonomy of follow-up queries and LLM classification techniques would further contribute to improving conversational search experience by supporting the personalization of information-seeking pathways and realistic simulation of user experience as well as a deeper understanding of user experience. 

% In this paper, we propose a taxonomy of conversational search interactions and utilize an LLM-powered classifier to support scalable and efficient evaluation of user satisfaction, along with interaction patterns. With our proposed approach, we demystify the user's purposes and actions of follow-up queries from the conversational search logs and relate them with user satisfaction scores. Through this investigation, we provide initial insights into understanding conversational search experience and enhance the process of user satisfaction evaluation. 

This paper makes the following contributions:
\begin{itemize}
    \item A taxonomy of users' follow-up query behaviors with conversational search capturing (1) users' motivations behind continuing conversations and (2) the patterns for follow-up queries.
    \item An LLM-powered classifier to automatically classify follow-up query patterns at scale
    \item Initial findings on how the distribution of follow-up query patterns differs by the levels of search satisfaction.
\end{itemize}

\section{Taxonomy Construction Process}

We describe our in-lab and real-world data collection and analysis process for building the taxonomy. 
Then, we explain the iterative process of concretizing the taxonomy with experimental data and LLM classification. 

\subsection{Step 1: Data Collection}
\label{sec:real-world}

\emph{Lab Study Setting:} 
We first collected conversational logs from an in-lab user testing of Naver Cue: with 9 predefined search tasks (Table~\ref{tab:experimental-tasks}): five exploratory (e.g., `Buy a dishwasher with a budget around \$500.') and four close-ended tasks (e.g., `Find how to get a government-subsidized housing loan.'). We selected these search tasks to include diverse task types, including searching for a location, shopping, living information, event planning, etc. 
All authors, including both academic researchers and research scientists from Naver, decided on the list of queries considering whether the query is realistic and frequently used by real-world users.
%Since a number are research scientists from the search engine company, we decided on the final nine tasks by discussing whether the query is realistic and frequently used by real-world users.
% All nine queries used in our data collection are shown in Table~\ref{tab:experimental-tasks}.
% \usepackage{booktabs}
\begin{table}[]
\caption{The list of tasks for the in-lab data collection.}
\label{tab:experimental-tasks}
{\footnotesize\begin{tabularx}{\columnwidth}{@{}lX@{}}
\toprule
Task No. & Task Description                                               \\ \midrule
E1          & Plan an overseas trip with parents                             \\
E2          & Choose a housewarming gift for a newly-wed couple              \\
E3          & Buy a dishwasher with the budget around \$500.                 \\
E4          & Get a exercise recommendation to train your back               \\
E5          & Choose a restaurant for a birthday party                       \\
C1          & Find the recent information on attempted Russian coup          \\
C2          & Find how to get a government-subsidized housing loan           \\
C3          & Find how to remove abrasive from a new stainless steel tumbler \\
C4          & Find whether it is okay to have painkiller for hangover        \\ \bottomrule
\end{tabularx}}
\vspace{-0.25cm}

\end{table}

We recruited participants with prior experience using generative AI applications (e.g., ChatGPT) but not generative search (e.g., Bing Chat, Bard), preventing their prior experience from influencing the quality of collected data. 
A total of 72 participants (36 Male, 36 Female; Age range from 18 to 64) conducted 108 search sessions. Half of the pool (36 participants) completed one conversation, and the other 36 completed two conversation tasks, with 5 minutes per task. Each participant was randomly assigned to one or two of the nine pre-defined tasks. 
% For participants who completed two conversations, we asked them to complete a self-reported survey on the user experience and the satisfaction of each conversation in a five-point Likert scale. 
Due to the system error, we lost the log of six participants who completed only one conversation task. 
We also excluded two single-turn sessions, and finally, we used 100 sessions in total.

% We used the conversation logs containing users' questions and the engine's response collected from two settings: (1) experimental and (2) in-the-wild.

% For the data collection, we first collected conversational logs from an experimental setting with a fixed set of tasks to limit the variety of conversational patterns for initial codebook construction. 
% % In the experimental setting, we used the conversational logs of 108 search sessions conducted by 72 participants in total. 
% Each participant received one or two search tasks to accomplish with a beta version of the conversational search engine. 
% We used nine search tasks to limit the variety of conversational topics: five exploratory tasks (E1 -- E5) and four close-ended tasks (C1 -- C4). 

% \input{tables/tasks}

\emph{Real-world Log Collection:} 
For real-world data, we randomly sampled 2,300 conversation tuples (the set of ‘User Query 1 - System Response - User Query 2’) from the live service of Naver Cue:, between January 1st -- February 17th, 2024. 
Following the company's internal offline evaluation protocol, we recruited eight external evaluators trained for the search evaluation. The evaluators rated the ``satisfaction'' of the search engine response on a 5-point Likert scale, defined by whether the search engine's response was satisfactory to meet the search intent of the query. 
The final satisfaction score was determined considering the following sub-criteria. 

\begin{itemize}
    \item \emph{Relevance}: Does the response understand the intent of the user query well? (0/1)
    \item \emph{Informativeness}: Does the response contain enough information to satisfy the search intent? (0/1/2)
    \item \emph{Trustworthiness}: Is the response factually correct? (0/1)
    \item \emph{Expressiveness}: Does the response sound natural without any awkwardness? (0/1)
    \item \emph{Safety}: Does the response contain sensitive or socially controversial content? (0/1)
\end{itemize}

Based on the rating results, we excluded 239 tuples where the external evaluators could not infer the search intent from the query or they rated that the engine completely failed to satisfy the search intent, resulting in a total of 2,061 tuples.

\subsection{Step 2: Iterative Taxonomy Creation with Qualitative Analysis}
%% with Lab-study Collected Data}
\label{sec:taxonomy-creation}
% We segmented each conversation from the in-lab conversation logs to produce conversational tuples consisting of n-th user query ($Q_n$), the search engine's response ($R_n$), and the follow-up user query ($Q_{n+1}$) for every user query in the conversation, ending up with 607 tuples in total. 
We followed a standard thematic analysis process to build our taxonomy.
We segmented each conversation from the in-lab conversation logs to produce conversational tuples, ending up with 607 tuples in total. 
Two authors independently conducted initial open coding on a total of 100 randomly sampled tuples and discussed together to produce an initial taxonomy. During the process, we identified two distinct axes: (1) the purpose of a follow-up query and (2) the expression/method of a follow-up query.
To revise the taxonomy, three authors additionally coded 150 tuples and iterated on the initial taxonomy. 
After the final revision of the taxonomy, we decided on the final codes for 250 tuples. 

\section{Taxonomy of Follow-up Queries in Conversational Search}

We describe our taxonomy across two major axes: (1) users' motivations behind continuing conversations (7 themes, Table~\ref{tab:axis1}) and (2) the patterns of follow-up queries (11 themes, Table~\ref{tab:axis2}).
% The themes are shown in Table~\ref{tab:axis1} and Table~\ref{tab:axis2}.
\begin{table*}[t]
\centering
\caption{Purpose of Follow-up Queries (Axis 1). Themes specific to conversational search are denoted with an asterisk (*). Relevant literature from the search domain is provided in the right column.}
\label{tab:axis1}

{\footnotesize\begin{tblr}{
  width=\textwidth,
  colspec={X[2.7]|X[2]|X[9]|X[1.8]},
  row{1} = {c},
  cell{3}{1} = {r=2}{},
  cell{7}{1} = {r=2}{},
  vline{2-3} = {1-9}{},
  vline{3} = {-}{},
  vline{4} = {-}{},
  hline{1-2,3,5-7,9,10} = {-}{},
  hline{4,8} = {-}{Silver},
}
\textbf{Category}                       & \textbf{Theme}                & \textbf{Definition}               & \textbf{Literature}               \\
\textbf{Clarifying Queries}          & {} & {When the user clarifies their search intent with the follow-up query.} & {\cite{rha2017exploration, sahib2012comparative, jansen2007Patterns}} \\
{\textbf{Broadening}\\\textbf{Information}}      & {Exploring\\ Domain}          & {When users seek a broad understanding of a topic and ask questions in order to accomplish the same task.} & {\cite{rha2017exploration}} \\
{}    & {Understanding Response}      & {When the user finds the response lacking important information or hard to understand and tries to gather more information.} & {\cite{rha2017exploration, athukorala2016exploratory}}\\
{\textbf{Deepening}\\\textbf{Information}}      & Narrowing Down                & {When the user seeks to obtain detailed and specific answers through the follow-up query.}  & {\cite{rha2017exploration}}    \\
{\textbf{Seeking}\\\textbf{Different}\\\textbf{Representations}\mbox{*}} & & {When the user seeks different representation (e.g. modality, format) to better understand the provided information } & {\cite{jansen2009Transition}}  \\

{\textbf{Evaluating}\\\textbf{Search Results}}  & {Verifying\\Information\mbox{*}} & {When the user requests additional explanations, alternatives, or evidence regarding the accuracy, completeness, or applicability of the information from the response.} & {-} \\
{} & {Reacting to\\Response\mbox{*}} & {When the user expresses the emotions of satisfaction or dissatisfaction or open ended feedback regarding the information provided by the response }  & {\cite{liao2020Conversational}} \\
\end{tblr}}

\end{table*}

\begin{table*}[t]
\centering
\caption{Actions of follow-up queries (Axis 2). Themes specific to conversational search are denoted with an asterisk (*). Relevant literature from the search domain is provided in the right column.}
\label{tab:axis2}

{\footnotesize\begin{tblr}{
  width=\textwidth,
  colspec={X[2]|X[3]|X[7]|X[2]},
  % cells={valign=m},
  row{1} = {c},
  cell{2}{1} = {r=3}{c},
  cell{5}{1} = {r=5}{c},
  cell{10}{1} = {r=2}{c},
  cell{12}{1} = {r=2}{c},
  % vline{1} = {-}{}
  vline{2-3} = {1-13}{},
  vline{4} = {1-13}{},
  vline{3} = {3-4,6-8,10-11}{},
  hline{1-2,5,10,12} = {-}{},
  hline{3-4,6-9,11,13} = {2-4}{Silver},
}
\textbf{Category}                          & \textbf{Theme}                & \textbf{Definition}                                                                       &\textbf{Literature}  \\
{\textbf{Query-}\\\textbf{specific}}    & Excluding Condition               & {When the user explicitly removes specific conditions from a previous query.}             & {\cite{rha2016Exploringa, rieh2006Analysis, jiang2013users}}                                                     \\
                                        &{ Adding/Specifying Condition}       & {When the user specifies additional conditions or parameters to include in the query.}     & {\cite{rha2016Exploringa, rieh2006Analysis, jiang2013users}}                                                    \\
                                        & Substituting Condition            & {When the user expresses the same query intent using different conditions or forms.}        & {\cite{rha2016Exploringa, rieh2006Analysis, jiang2013users}}  \\
{\textbf{Response-}\\\textbf{specific}} & Converting Format\mbox{*}         & {When the user requests to receive the answer in a specific format (e.g., table, map, graph).}   & \cite{wang2009Broad}                                              \\
                                        & Criticizing Response\mbox{*}      & {When the user points out that the provided answer is unrelated or inappropriate to the original question.}  & {\cite{liao2020Conversational}} \\
                                        & Affirming Response\mbox{*}         & When the user expresses satisfaction with the provided answer.  & {\cite{liao2020Conversational}}  \\
                                        & {Requesting Additional Information} & When the user requests additional information related to the response.  & {\cite{rha2017exploration, athukorala2016exploratory}}  \\
                                        & Confirming Response                 & {When the user directly requests reconfirmation of the user's understanding or verifies previous response.} & {\cite{ashktorab2019resilient, cambazoglu2021intent}} \\
{\textbf{Session-}\\\textbf{specific}} & {Requesting Related Information}     & {When the user requests related information that is not directly referred to in the response but related in the high-level domain}  & {\cite{boldi2011query}}                                                      \\
                                        & Requesting Opinion\mbox{*}          & {When the user asks additional information or opinion on the users’ own alternative options not previously referred to in response.} & {\cite{cambazoglu2021intent}}                                 \\
\textbf{Misc.}                          & Chatting Casually\mbox{*}          & When the user attempts chitchat rather than seeking specific information. & {\cite{liao2020Conversational}} \\
                                        & Requesting Unrelated Information   & When the intent or content of the question is unclear.                    & {-} \\
\hline
\end{tblr}}
\end{table*}

\subsection{Axis 1: User Motivation}
The first axis covers the motivations behind the follow-up queries to continue the conversation. Previous work on the conceptual framework of conversational search~\cite{sahib2012comparative, liao2020Conversational} defined three stages of the information-seeking process:\textit{query formulation}, \textit{search results exploration}, and \textit{query reformulation}.
% ~\cite{sahib2012comparative} and connected it to a conceptual framework illustrating the intent of search users. 
Our themes show how \textit{query reformulation} and \textit{search results exploration} stages are conducted with LLM-powered conversational search.

\textbf{Clarifying queries}, indicating users' attempts to re-express their search intent in different words, has been recognized as `query reformulation' in not only traditional search~\cite{rha2017exploration} but also in conversational search contexts~\cite{Azzopardi2018, liao2020Conversational}. 
Clarifying queries shows the process of users refining the queries to align their intent with the system's capabilities~\cite{Zamani2020Generating}. 
% but also enhance user satisfaction by facilitating a deeper and more effective exploration of conversational search~\cite{Aliannejadi2019Asking}.
% Building from previous literature, these clarifying queries not only refine the search process by accurately aligning user intent with system capabilities~\cite{tavakoli2020generating, Zamani2020Generating} but also enhance user satisfaction by facilitating a deeper and more effective exploration of conversational search~\cite{Aliannejadi2019Asking}.

In contrast, \textbf{Broadening Information}, \textbf{Deepening Information}, and \textbf{Seeking Different Representations} reveal diverse expressions of users' endeavors to navigate the relevant information space, expanding result exploration behaviors presented in previous work~\cite{Azzopardi2018, liao2020Conversational}.
Although those motivations have also been identified in prior research on traditional search~\cite{rieh2006Analysis, jansen2007Patterns}, we observed several themes became more explicit in conversational search. 
\textbf{Seeking Different Representations} was one case that became explicit in the queries of conversational setting, which involved queries asking for the search results in different or complex forms like photos and tables (e.g., ``Show the posture for wall squat in photos (E4)'', ``Compare the strength and weakness of the previous search results in a table (E3)''). With conventional search engines, such intentions would have been expressed as switching to image search or browsing and sensemaking multiple web pages rather than separate queries.
% (e.g., exploring different sources such as images, maps, etc.).  
% \textbf{Changing Modalities} became explicit in the queries of conversational setting, which would be expressed as a switch of search engines (e.g., document search to image search) or users' internal sensemaking from browsing the results (e.g., tabularizing multiple options) with traditional search engines. 

We also observed that users are \textbf{Evaluating Search Results} by seeking confirmations and providing feedback on the results to improve the search quality, distinguishing it from other themes where information exploration dominated. This behavior aligns with interrogating behaviors from conversational searches proposed by Azzopardi et al.~\cite{Azzopardi2018, liao2020Conversational}, but not previously observed in conventional search.

The new themes, including \textbf{Verifying Information}, \textbf{Reacting to Response}, as well as \textbf{Seeking Different Representations}, show more diverse expectations for conversational search engines, such as verifying the trustworthiness of its response, gathering multiple pieces of information for their sensemaking process, or understand users' search intentions better based on their natural language feedback. 
Such new themes of search intentions suggest that people had a new mental model for search engines. With conventional search engines, users considered that the search engine's capability would be limited to retrieving documents relevant to the query. However, with the extensive capabilities of LLM-based search, users would have a new mental model that the search engines would be capable of handling more complex and implicit queries. 
% 사람들이 model 의 성능을 높이고자하는 마음은 계속 있었지만, conversational query 가 가능해지자 그런 목적으로 사용하기 시작했다 
% 이거야말로 쿼리 자체의 자유도가 확 높아지면서 사람들의 멘탈모델이 달라져서 이런 evaluation 목적으로 query 를 사용하기 시작한 경우.
% 기존의 search 에도 keyword based 였기 때문에, 자연어가 아닌 것은 아니었지만 (자유로운 표현이 어려운 것은 아니었지만), 사람들이 그러한 목적으로 쿼리를 사용하지 않으며, 오히려 검색 최적화 (?) 에 관심을 보였다는 점이 큰 차이점이라는 생각.

% These new themes imply that users engage in conversational interactions not just to refine their search intents but also to expect the engine to play additional roles, such as collaborating in the decision-making process.
% 추가적인 역할을 기대한다는 것으로도 볼 수 있음. 기존에 서치엔진이 할 수 없다고 생각했던 인터랙션이 가능해져서 새로운 양상의 인터랙션이 일어난다~ : discussion 

\subsection{Axis 2: Patterns of Follow-up Queries}
The second axis represents the actions users exhibit in their follow-up queries. We classify them into three categories: \textbf{Query-specific}, \textbf{Response-specific}, and \textbf{Session-specific}. 
% defined in terms of previous queries, responses, and the overall conversation context. 
% , similar to patterns of query reformulations with traditional searches~\cite{rieh2006Analysis}.
\textbf{Query-specific} follow-up queries involve relatively simple and well-known refinements to the original query, such as \textbf{Excluding Condition}, \textbf{Adding/Specifying Condition}, and \textbf{Substituting Condition}~\cite{rieh2006Analysis, jiang2013users, trippas2018Informing, rha2016Exploringa, wang2009Broad}. 
% These modifications are commonly observed in both traditional~\cite{rieh2006Analysis} and conversational search~\cite{trippas2018Informing}.
% , alongside \textbf{Response-specific} queries. 
\textbf{Response-specific} follow-up queries, such as \textbf{Converting Format} and \textbf{Requesting Additional Information}, show how browsing search results with a traditional search interface translate to conversational actions.
\textbf{Criticizing Response} or \textbf{Affirming Response} on the results highlight a distinctive aspect of conversational searches with naturalistic interactions. 
% We structured the second axis to prioritize the classification of follow-up actions, starting with query-specific actions. If an action does not qualify as query-specific, we then evaluate if it can be categorized as response-specific. If it also does not fit into the response-specific category, only then do we consider it as session-specific. 
These themes break down the category of \textit{Conversational Feedback} identified in conversational QA settings~\cite{liao2020Conversational}.
% into more detailed components, thereby enhancing the understanding of conversational search feedback. 
Users' actions of \textbf{Confirming Response} of the search engines involve double-checking the engine's responses, similar to confirmation actions of the conversational agents, commonly provided as a repairing strategy for overcoming conversational breakdowns~\cite{ashktorab2019resilient}.
% toward the search engine, which can be easily observed as a repairing strategy of chatbots to overcome conversational breakdowns~\cite{ashktorab2019resilient}. Contrary to systems' usage of confirming response to acknowledge the breakdown, users directly requests verification on previously provided answers or information.

From a broader perspective, users sometimes attempt to search for information to achieve their holistic search tasks, such as decision-making based on the information found, with \textbf{Session-specific Queries}.
% While the existing taxonomy of query reformulation coarsely grouped those kinds of queries as \textit{parallel move}~\cite{boldi2011query}, characterized by lateral topic shifts in query reformulation patterns, signals from conversational search queries enable unpacking \textbf{Session-specific} patterns. 
For instance, \textbf{Requesting Opinions} to the system reflects users' attempt to actively ask for opinions and perspectives on the users' alternative ideas or options to complete their search task.
\textbf{Requesting Related Information} presents the users' intent to seek information related to their high-level search task but not directly referred to in the search results. 

\section{Classifying Interaction Patterns with LLM}

To apply our taxonomy to classify conversational logs at scale, we built an LLM-powered classifier to analyze interaction patterns of diverse conversational search logs. 

\subsection{Implementation and Iteration of LLM-based Classifier}
%조금 많이 자세하게 썼습니다.

We iteratively designed an LLM prompt for GPT-4 (gpt-4-1106-preview, temperature = 1) to automatically code each conversational tuple with our taxonomy, following existing approaches on qualitative coding with LLMs~\cite{chew2023llm, xiao2023supporting}. We implemented the LLM-based classifier with three considerations: coverage, accuracy, and efficiency.
We first designed the initial version of the classifier using the definitions of each theme. We tested it with 250 manually coded conversational tuples from the taxonomy generation process (Section~\ref{sec:taxonomy-creation}).
Below, we describe how we iterated our classifier in terms of each consideration.

\emph{Coverage}: To ensure comprehensive coverage of conversational elements, we added the ``Unclassified'' theme to the classifier to identify queries that did not correspond with any of our predefined themes. We examined an additional 1,700 conversational tuples from in-the-wild conversational logs and newly identified tuples not captured by the taxonomy. Based on the results, we incorporated additional themes, such as casual chatting for Axis 2, to complement our taxonomy.

\emph{Accuracy}: To enhance the accuracy of our taxonomy classification, we first identified the conversational tuples that were frequently misclassified or inconsistently classified across multiple runs.
To mitigate them, we refined the wording of the taxonomy definitions to clarify ambiguous ones.
We also asked GPT-4 to explain the reasons behind its misclassification. For each reason, we selected one representative conversational tuple and included it in the classifier prompt as a few-shot example. 
Additionally, we implemented a priority-based decision mechanism in the classifier, instructing it to prioritize \textbf{Query-Specific} classifications over \textbf{Response-Specific}, followed by \textbf{Session-Specific}, and finally \textbf{Miscellaneous} taxonomies. This was to ensure that the classifier consistently outputs the most relevant taxonomy in cases of ambiguous classifications.
Given that our datasets were in Korean, we also compared the performance of English versus Korean prompts. The English prompts consistently yielded higher accuracy, highlighting a significant increase in performance between the two languages. 

\emph{Efficiency}: For the final prompt iteration of our classifier, we focused on optimizing the prompt length while preserving the accuracy. This was to create a scalable classifier suitable for large-scale analyses. Among the few-shot examples of conversational tuples in the prompts, we removed search engines' lengthy responses that did not help identify the intent (Axis 1) or the action (Axis 2).

After these refinements, we evaluated the performance of our classifier by comparing its accuracy with that of human labels. Human evaluation was done by two authors in our research team --- with the process of (1) individual labeling and (2) discussing the differences to solve the conflict. The tuple data used for performance evaluation was randomly chosen from a lab setting (N = 247), which was not used during the construction of the taxonomy and classifier. In the end, the classifier achieved an accuracy of 74.39\% for user motivations (Axis 1) and 71.11\% for the patterns of follow-up queries (Axis 2).

\subsection{Application of Taxonomy onto Real-World Conversation Logs}

\begin{figure}
    \centering
    \begin{subfigure}[b]{0.7\textwidth}
        \includegraphics[width=\columnwidth]{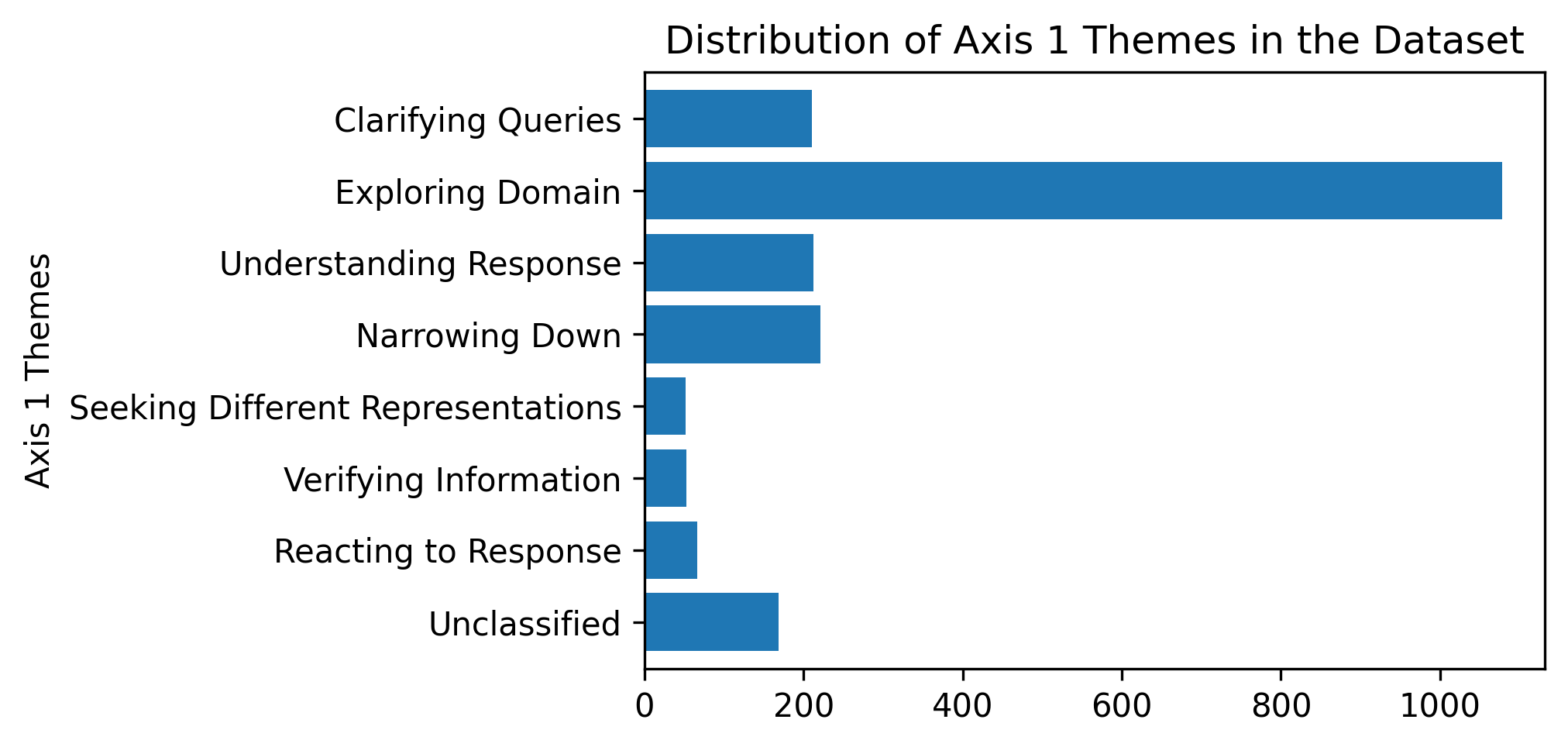}
    \end{subfigure}
    \begin{subfigure}[b]{0.7\textwidth}
        \includegraphics[width=\columnwidth]{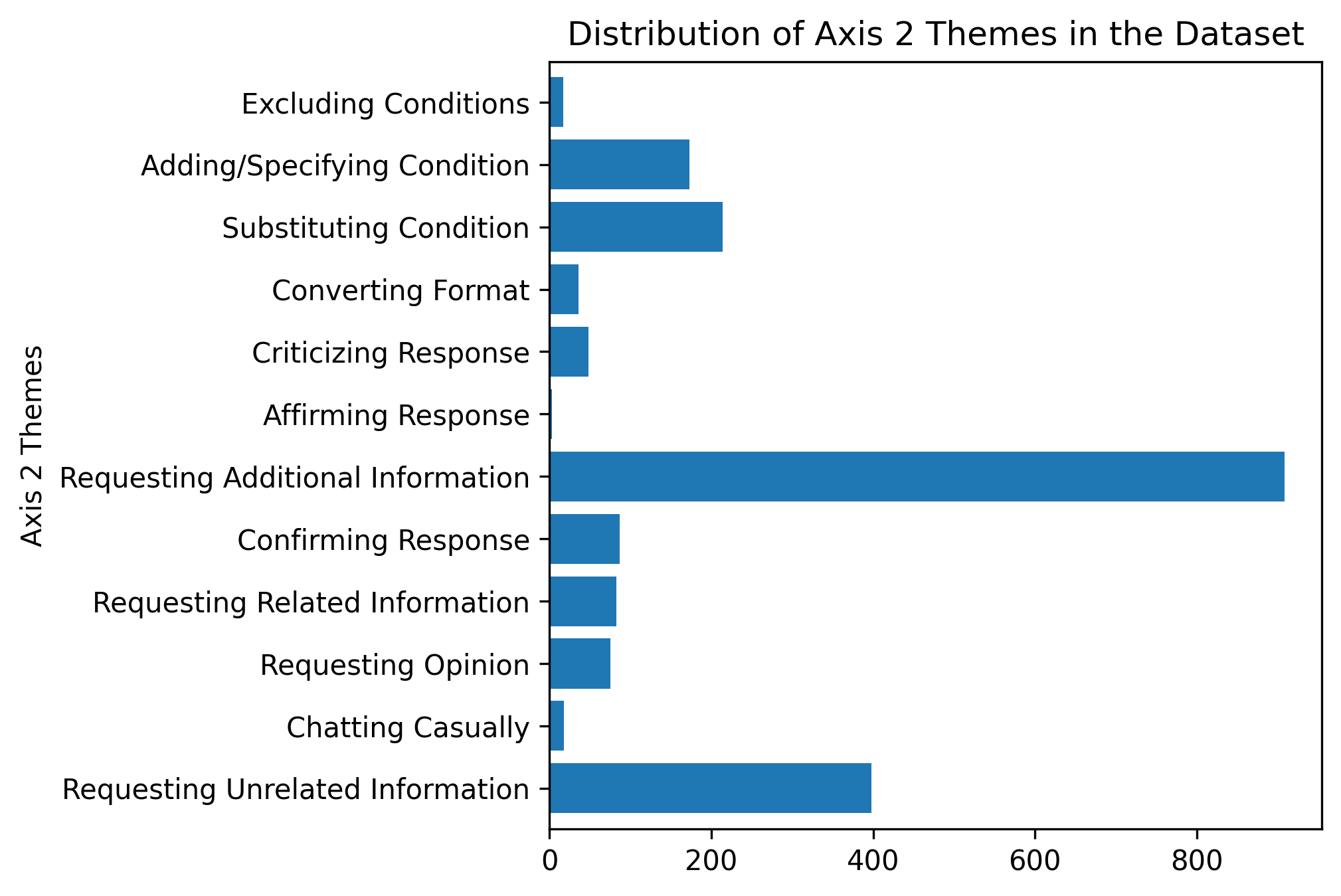}
    \end{subfigure}
    \caption{Frequency distribution of our taxonomy from the real-world data}
    \label{fig:real-world}
\end{figure}

% We applied our LLM-powered classifier to 300 randomly sampled sessions from real-world conversation logs containing 1,714 conversational tuples.
Using our LLM-powered classifier, we analyzed 2,061 conversational tuples randomly chosen from real-world conversation logs (Section~\ref{sec:real-world}). 
Our classifier classified the purposes (Axis 1) of more than 91\% of the tuples and the actions (Axis 2) of more than 80\% of the tuples into taxonomy, while the remaining tuples were left `unclassified.'

For Axis 1 (Purpose), the most frequent themes were \textbf{Exploring Domain}, followed by \textbf{Narrowing Down}, \textbf{Understanding Response}, and \textbf{Clarifying Queries}. 
For Axis 2 (Action), the most frequent themes were \textbf{Requesting Additional Information}, \textbf{Substituting Condition}, and \textbf{Adding/Specifying Conditions}. 

\begin{figure}
    \centering
    \includegraphics[width=0.7\columnwidth]{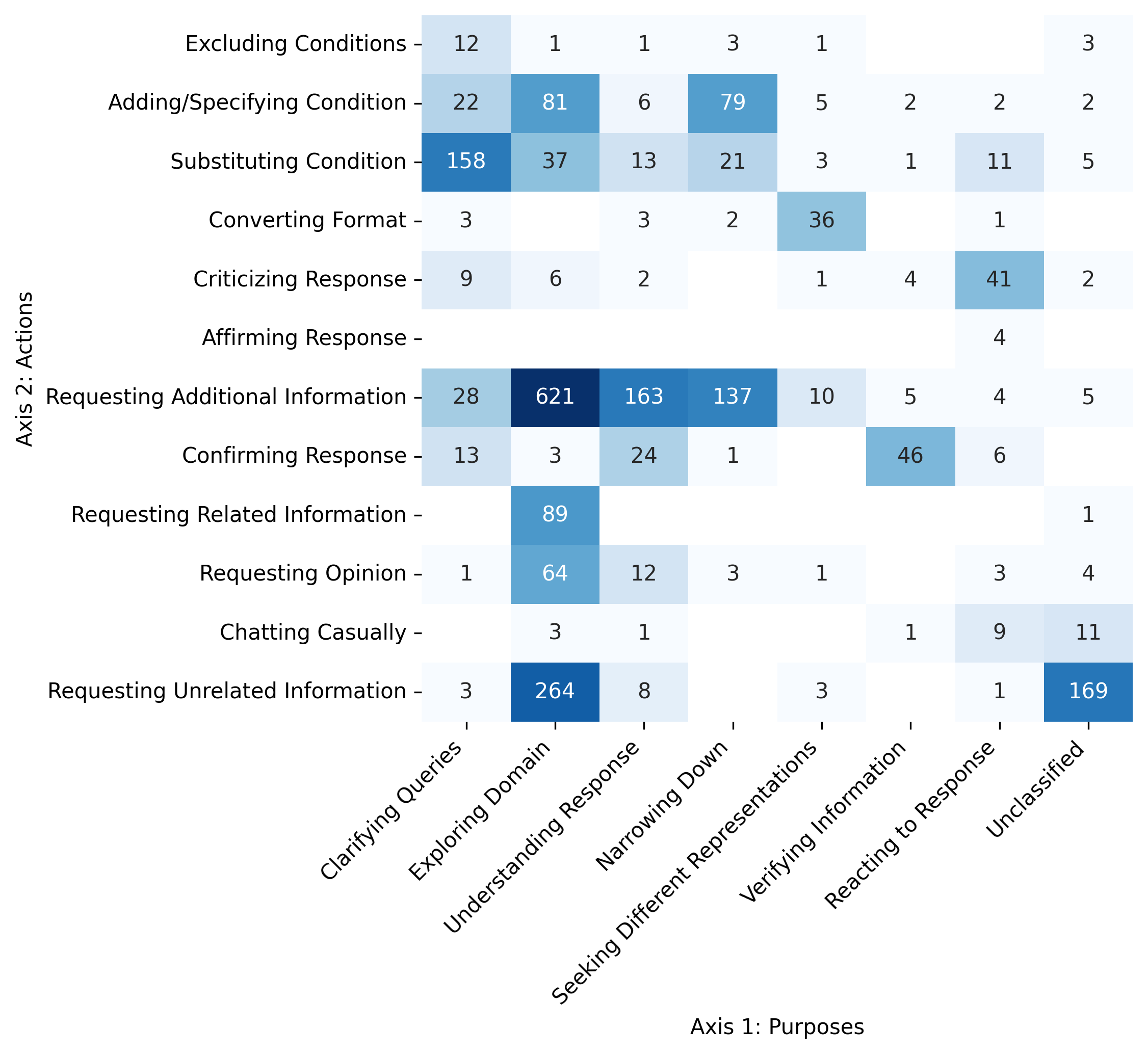}
    \caption{The co-occurrences of themes between Axes 1 and 2}
    \label{fig:cooccurrence}
\end{figure}

We examined the themes' co-occurrences to observe the correlation between the axes (Figure~\ref{fig:cooccurrence}). The co-occurrence suggested users' common actions to achieve their purpose, such as \textbf{Requesting Additional Information} and \textbf{Requesting Unrelated Information} for \textbf{Exploring Domain}, \textbf{Substituting Condition} for \textbf{Clarifying Queries}, \textbf{Requesting Additional Information} for \textbf{Understanding Response}, and \textbf{Adding/~Specifying Condition} for \textbf{Narrowing Down}. 

\section{Conversational Patterns vs. Perceived Search Experiences}

\begin{figure}[ht]
    \centering
    \begin{subfigure}[b]{\columnwidth}
        \includegraphics[width=\columnwidth]{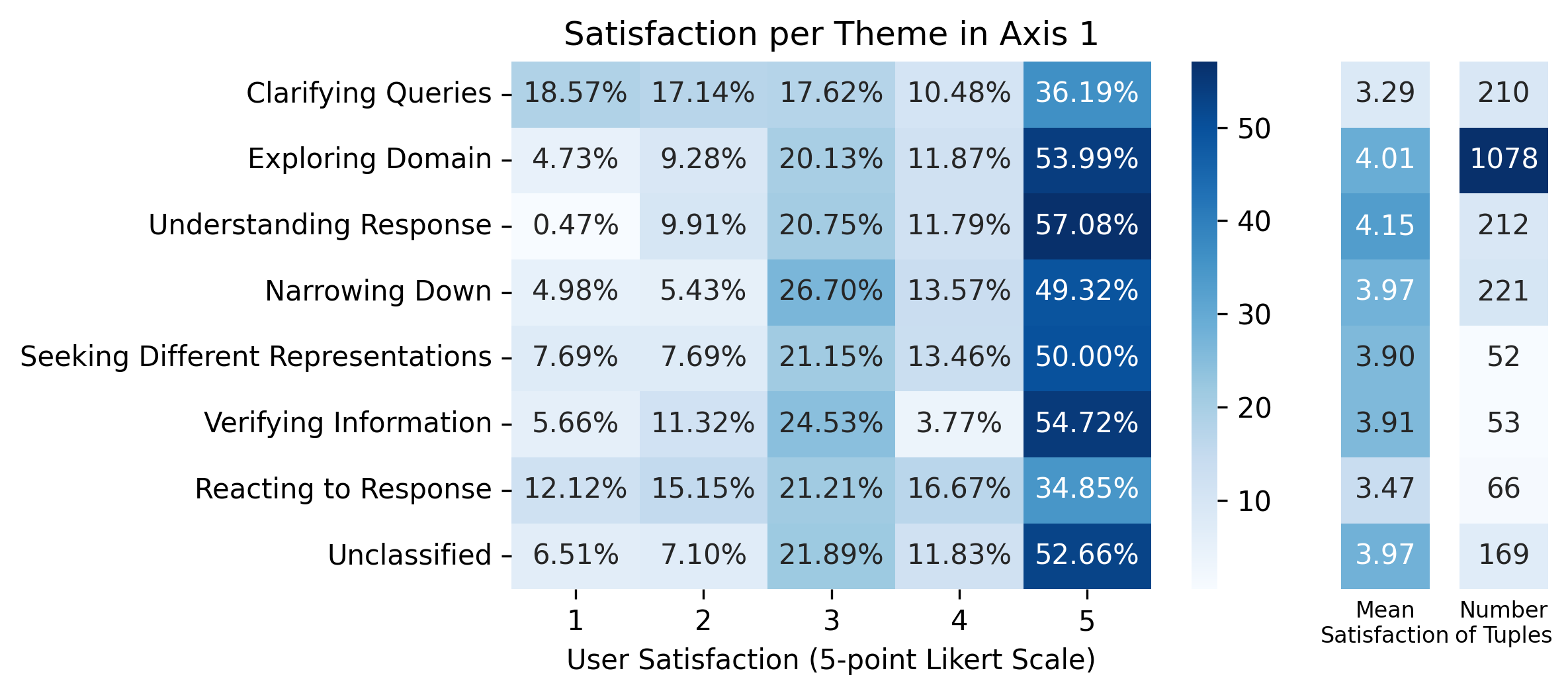}
    \end{subfigure}
    \begin{subfigure}[b]{\columnwidth}
        \includegraphics[width=\columnwidth]{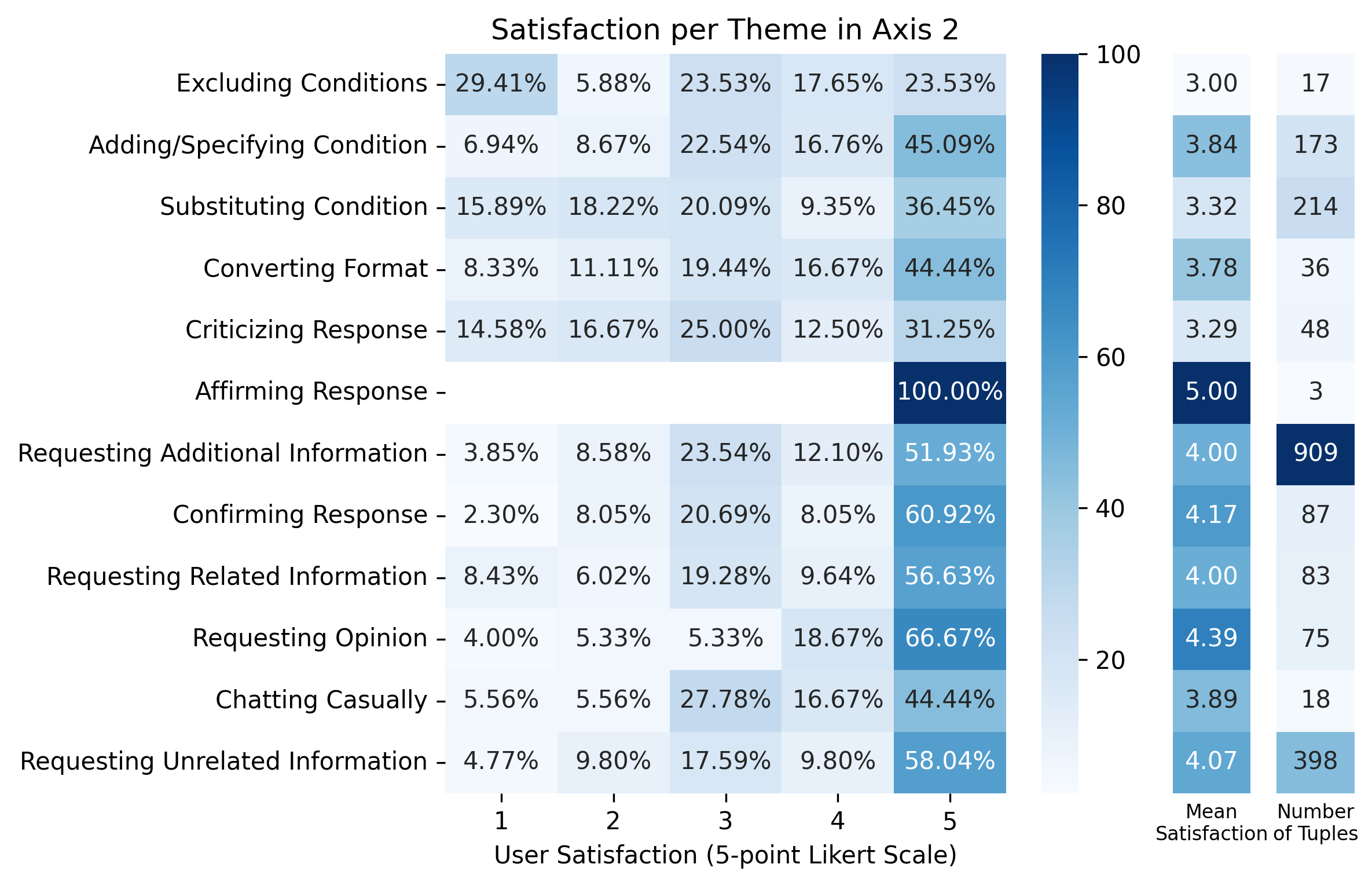}
    \end{subfigure}
    \caption{Distribution of satisfaction score by each theme.}
    \label{fig:satisfaction-distribution}
\end{figure}

\begin{figure}
    \centering

    \includegraphics[width=\columnwidth]{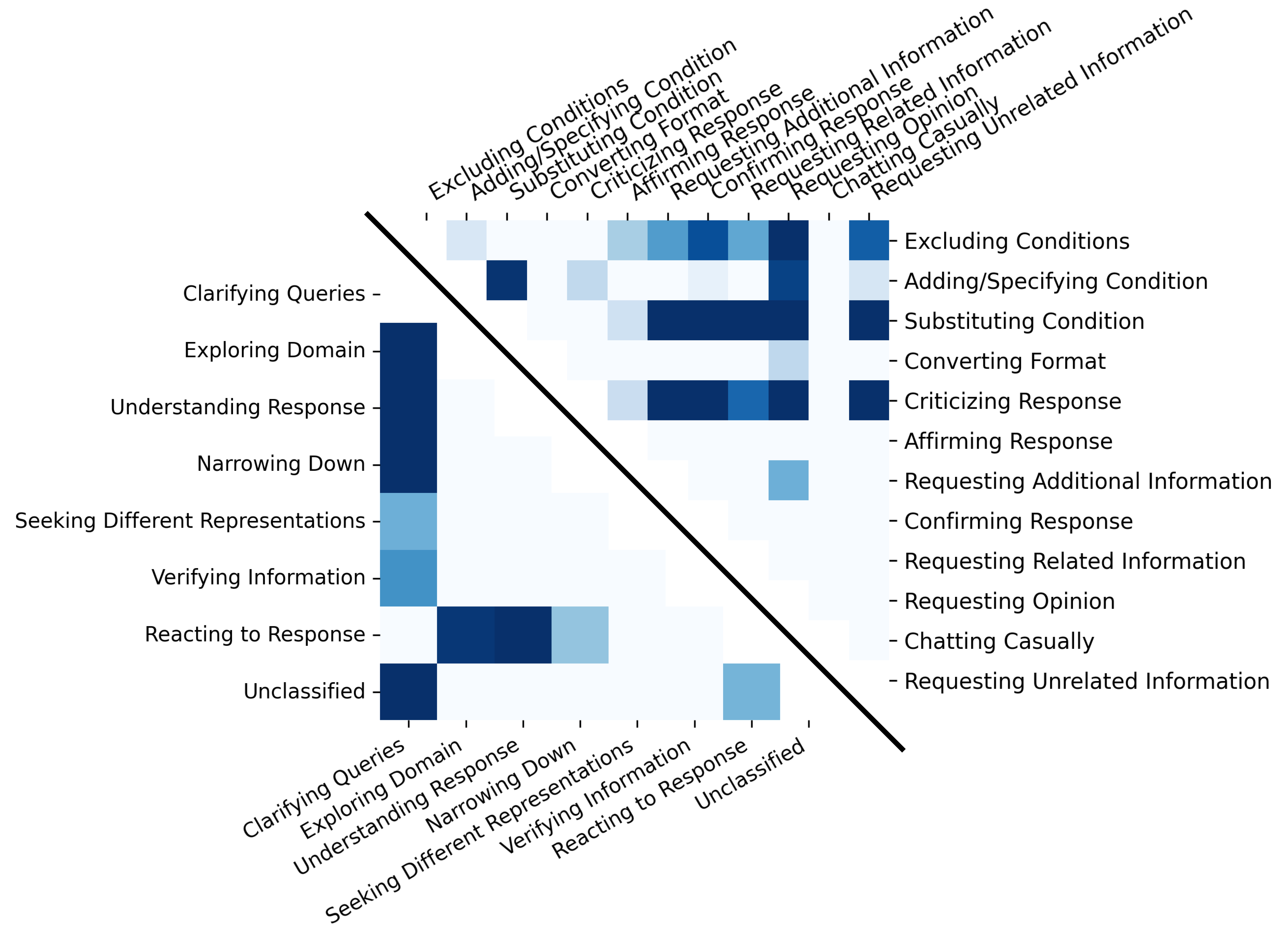}

    \caption{Result of pairwise comparison of satisfaction score by each theme. The colored squares signify statistically significant differences in satisfaction scores between the themes.}
    \label{fig:satisfaction-pairwise}
\end{figure}

\begin{figure}
    \centering
    \begin{subfigure}[b]{\columnwidth}
        \includegraphics[width=\columnwidth]{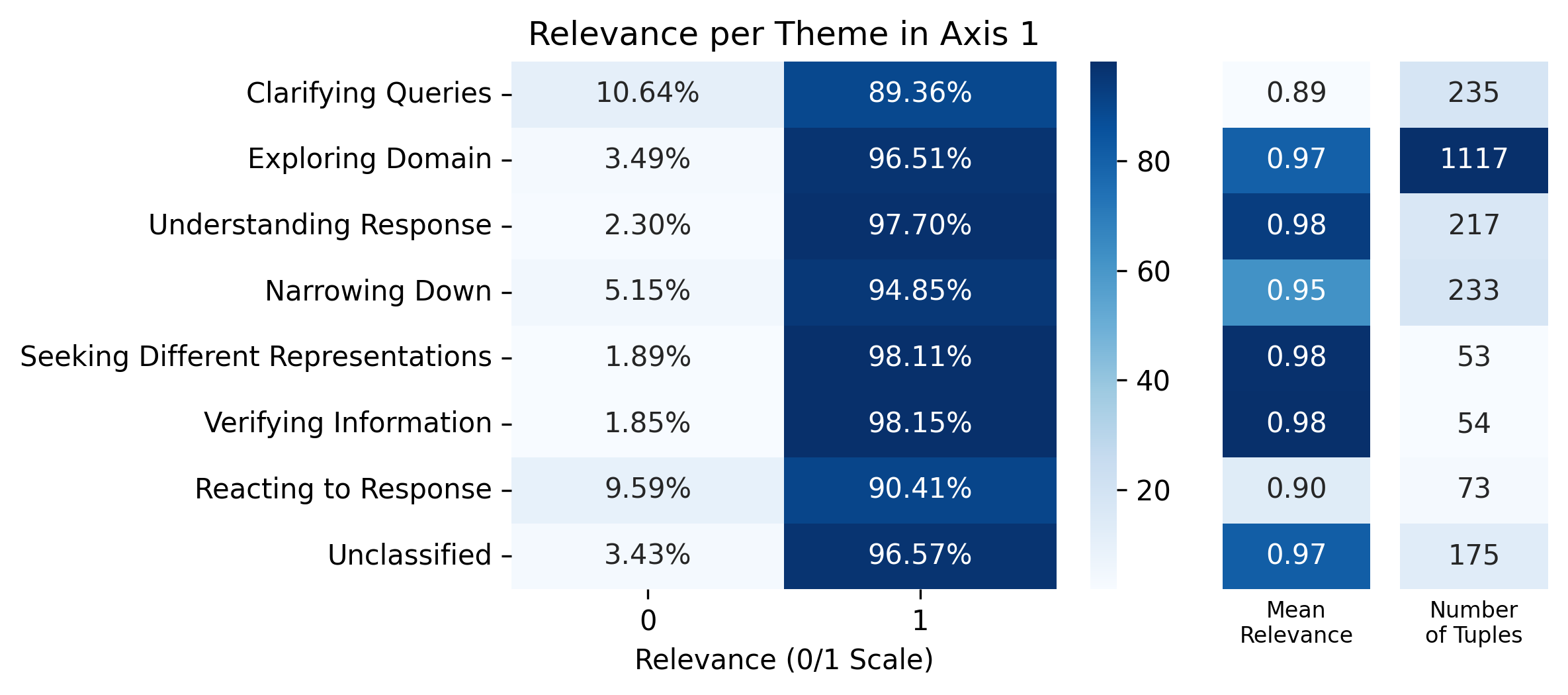}
    \end{subfigure}
    \begin{subfigure}[b]{\columnwidth}
        \includegraphics[width=\columnwidth]{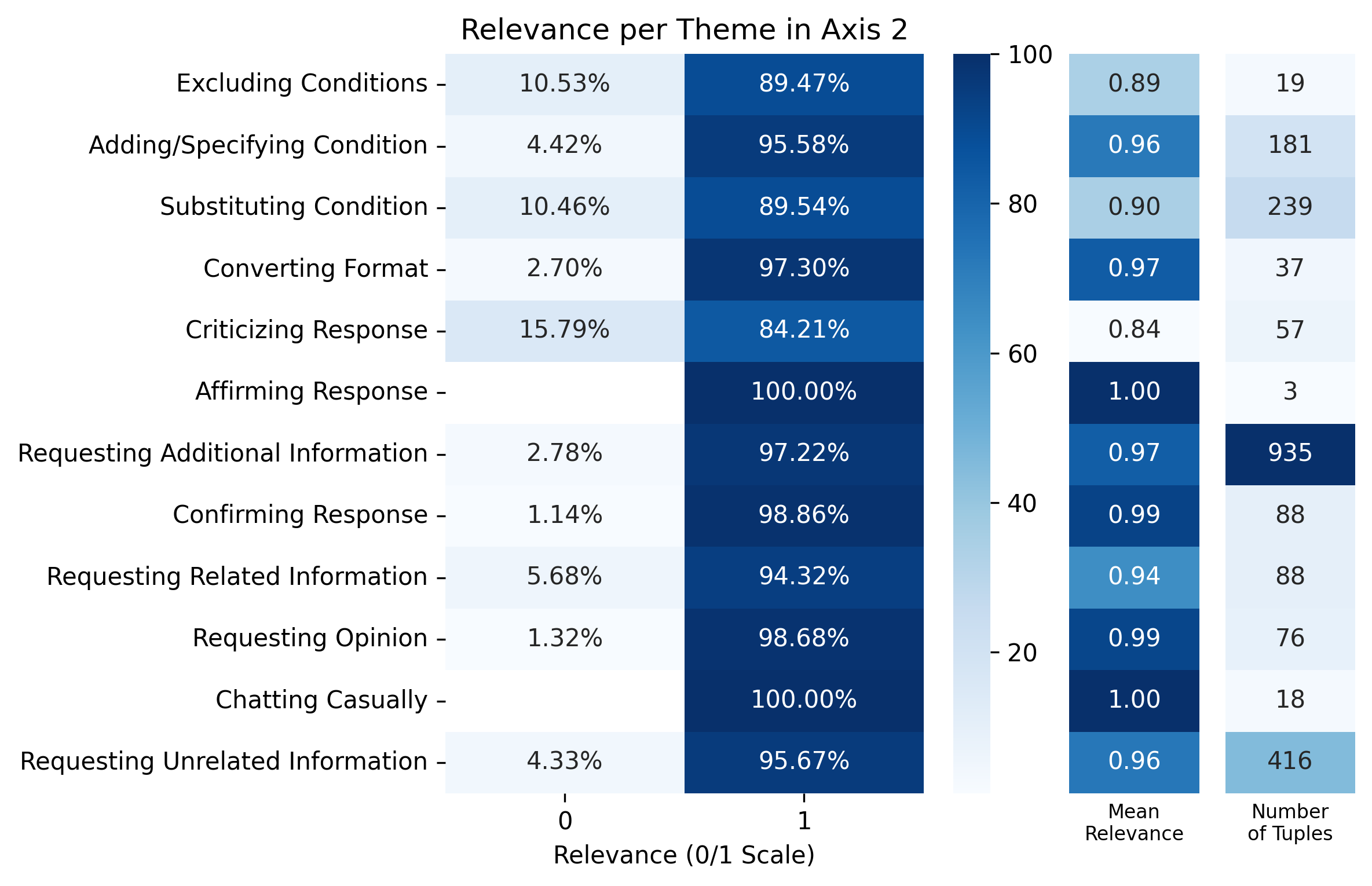}
    \end{subfigure}
    \caption{Distribution of relevance score by each theme.}
    \label{fig:relevance-distribution}
\end{figure}

\begin{figure}
    \centering
    \begin{subfigure}[b]{\columnwidth}
        \includegraphics[width=\columnwidth]{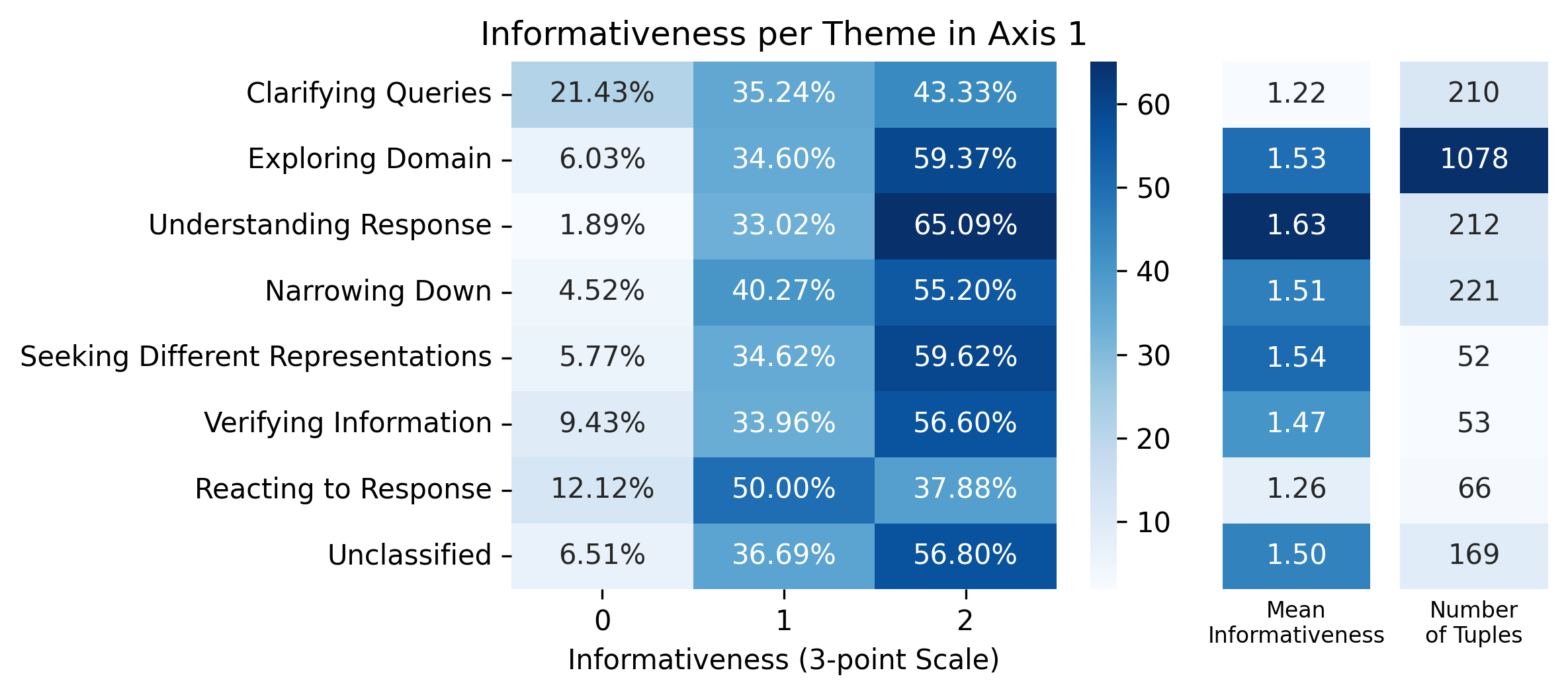}
    \end{subfigure}
    \begin{subfigure}[b]{\columnwidth}
        \includegraphics[width=\columnwidth]{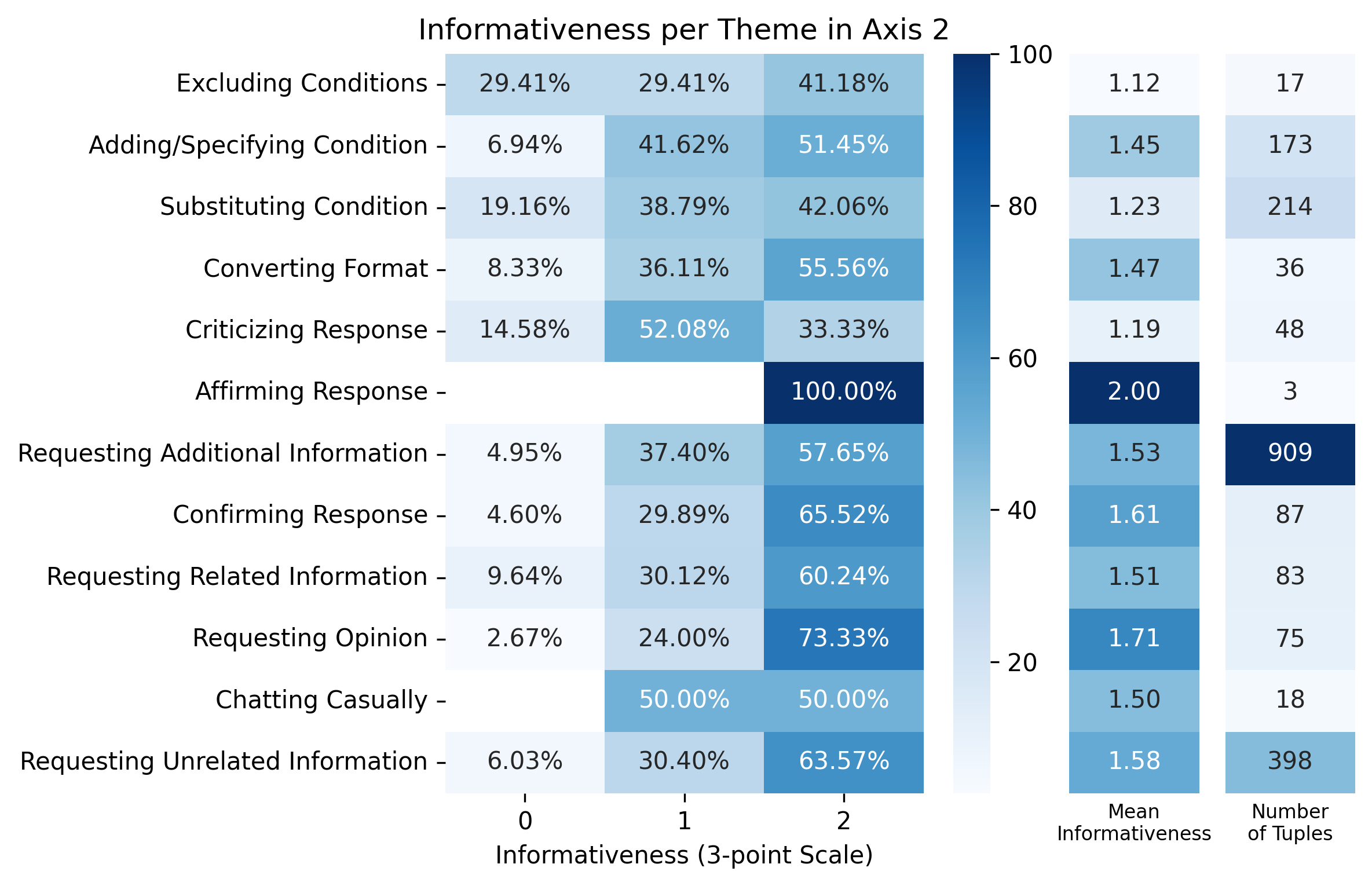}
    \end{subfigure}
    \caption{Distribution of informativeness score by each theme.}
    \label{fig:informativeness-distribution}
\end{figure}

\begin{figure}
    \centering
    \begin{subfigure}[b]{\columnwidth}
        \includegraphics[width=\columnwidth]{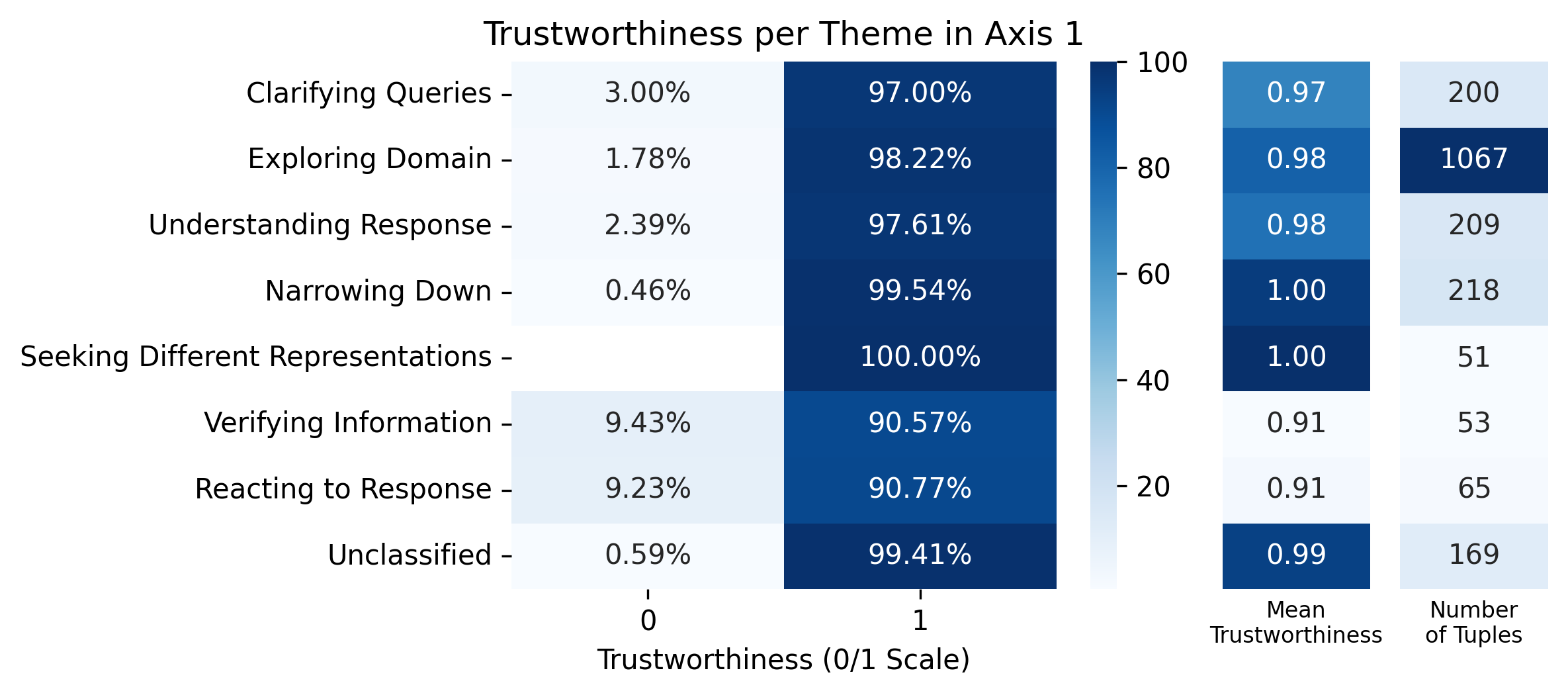}
    \end{subfigure}
    \begin{subfigure}[b]{\columnwidth}
        \includegraphics[width=\columnwidth]{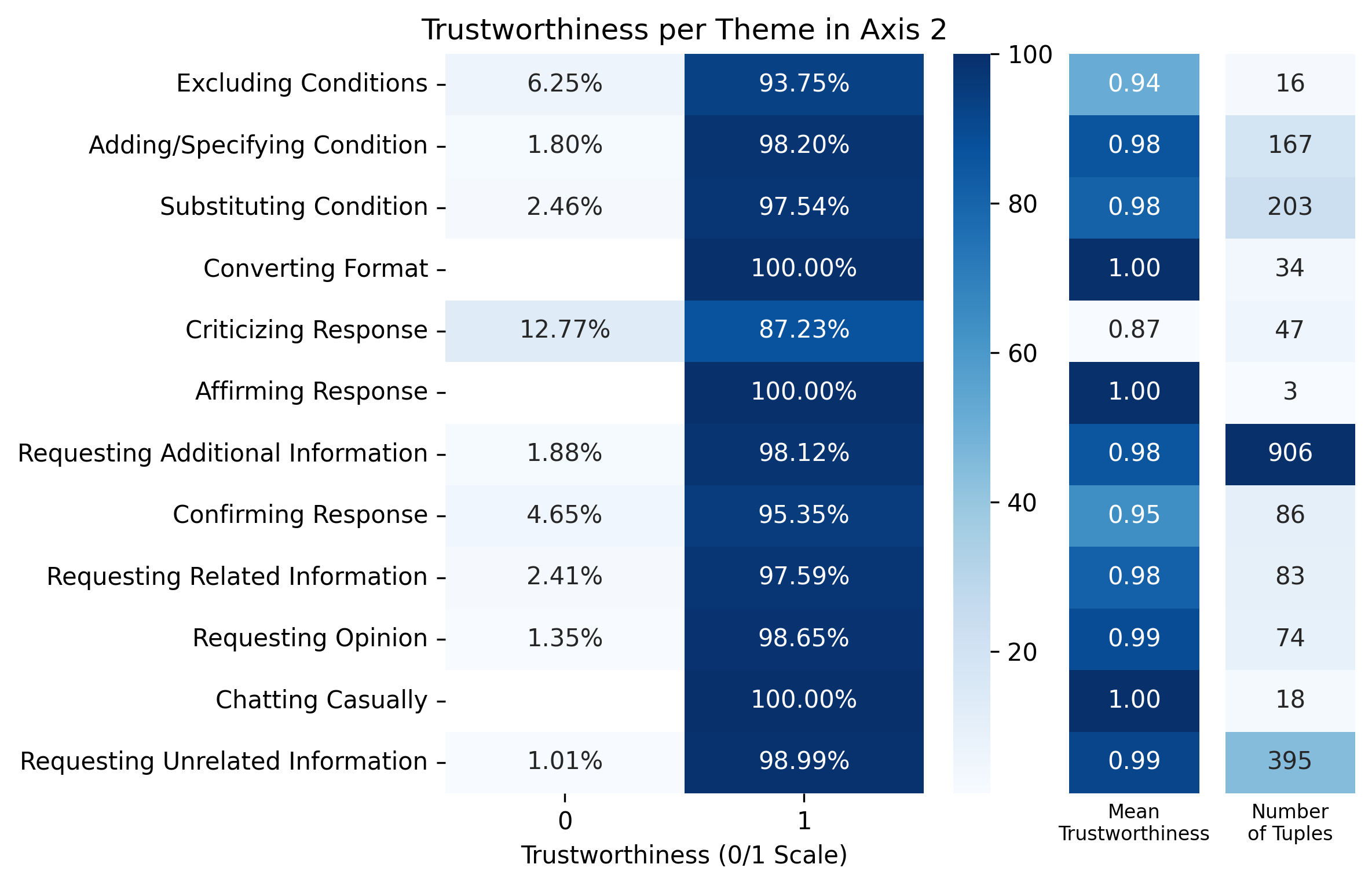}
    \end{subfigure}
    \caption{Distribution of trustworthiness score by each theme.}
    \label{fig:trustworthiness-distribution}
\end{figure}

% Existing work on query reformulation~\cite{hassan2013clicks} and conversational signals~\cite{liao2020Conversational} showed that users' query patterns signal satisfaction with the search experience. 
% Following existing work, we examined whether our themes could implicitly signal users' search experiences. 
With the classifier, we analyzed real-world conversation logs by themes in our taxonomy and observed how each theme correlates with users' search satisfaction.
The user satisfaction score of each theme in our taxonomy shows that some of the themes have a high correlation with user satisfaction (Figure~\ref{fig:satisfaction-distribution}).
% Juho Kim: this is a bold statement we don't have strong evidence for. let's just say some of the themes showed a high correlation with user satisfaction. Being able to use as signals is a bit speculative.

We first ran a Kruskal-Wallis test and verified that the satisfaction scores were significantly different by each theme (For Axis 1, H = 56.65, p $<$ 1e-9. For Axis 2, H = 79.88, p $<$ 1e-12). 
For Axis 1, the post-hoc analysis with the Dunn test between the themes showed that \textbf{Clarifying Queries} and \textbf{Reacting to Response} had significantly different satisfaction scores compared to other themes (Figure~\ref{fig:satisfaction-pairwise}).
The heatmap of themes and satisfaction scores (Figure~\ref{fig:satisfaction-distribution}) showed that \textbf{Clarifying Queries} and \textbf{Reacting to Response} had lower satisfaction scores than other themes. 
This initial observation shows that actively correcting prior queries and reacting to the system response can be potentially negative feedback given by the users. 
% This agrees with previous work on query reformulations in conventional search, suggesting that query reformulations signal unsatisfied user needs~\cite{hassan2013clicks}. 
Moreover, analysis of detailed metrics showed more nuanced features. 
In the case of \textbf{Clarifying Queries}, the responses tend to be rated with very low informativeness (Figure~\ref{fig:informativeness-distribution}), while the responses classified as \textbf{Reacting to Response} were rated low regarding credibility (Figure~\ref{fig:trustworthiness-distribution}).   
% On the other hand, \textbf{Understanding Response} and \textbf{Exploring Domain} had higher satisfaction scores, suggesting that users were satisfied with the given information and trying to move on to other search intents. 

For Axis 2, the post-hoc analysis with the Dunn test between the themes showed that \textbf{Excluding Condition}, \textbf{Substituting Condition}, and \textbf{Criticizing Response} had significantly different satisfaction scores from other themes. 
The heatmap of themes and satisfaction scores (Figure~\ref{fig:satisfaction-distribution}) showed that  \textbf{Excluding Condition}, \textbf{Substituting Condition}, and \textbf{Criticizing Response} were correlated with low satisfaction. The co-occurrence of the themes suggested that the first two themes were frequently used to clarify the users' search intents (Figure~\ref{fig:cooccurrence}). 
\textbf{Criticizing Response}. It may suggest that the users' dissatisfaction with the engine's response was related to the trustworthiness of the response (Figure~\ref{fig:trustworthiness-distribution}).

\section{Discussion \& Future Work}

% From the analysis, we discovered new patterns apparent from the conversational search settings but not directly present with traditional search. 
% In this work, we present patterns of follow-up queries with conversational search, which capture motivations or behaviors similar to or different from interactions with traditional search. 
We discuss the potential implications of our taxonomy, the potential challenges of the automatic evaluation, and the limitations of this work.

\subsection{Implications and Future Application of the Taxonomy}
We first investigated various types of follow-up queries in the context of conversational search. Expanding and comparing with previous taxonomies, our taxonomy could contribute to a better understanding of users' intent behind follow-up queries, thereby uncovering new types of user needs in conversational search settings. 

We expect that such understanding could be used to build search engines taking initiative of information seeking process to actively guide users. For example, if users issue a specific pattern of follow-up query (e.g., \textbf{Narrowing Down}) more often when looking for a specific type of information (e.g., local information), analyzing the actual queries could give insights on the specific information that the users frequently desire. Such insights could be embedded into the search system as query recommendations~\cite{baeza2004query} or clarifying queries from the search engine~\cite{Zamani2020Generating}. Furthermore, insights from the follow-up query patterns of individual users could support designing personalized search paths for each user.   

Also, our initial findings on the correlation between the users' behavioral patterns and satisfaction show how our taxonomy could guide improving the user experience with conversational search. 
For example, \textbf{Clarifying Queries} and \textbf{Substituting Condition} could directly relate to users' challenges in conveying their search intent to the engine. 
On the other hand, \textbf{Understanding Response} and \textbf{Requesting Additional Information} suggest that the users were satisfied with the response in general but needed more information to completely satisfy their search intent, which could inform directions for improving the engine performance.  
% Our analysis of user satisfaction and the distribution of themes suggests that some purposes and actions may have correlations with satisfactory search experience. 
Since we observed the relationships between axes and how user satisfaction relates to each theme within these axes, we can integrate these insights to identify signals of unsatisfactory sessions. 

We also envision that our taxonomy could aid a more realistic conversation simulation. As conversational search across multiple turns can create numerous amount of potential pathways of user interaction, we expect that user simulation would be essential to holistically examine and evaluate conversational search in a user-centric manner. 
Recent work has explored simulating information-seeking behaviors with conventional search engines~\cite{zhang2024usimagent, engelmann2024context, owoicho2023exploiting, sekulic2022evaluating}. 
At the same time, the HCI and AI communities have been actively investigating LLM-powered user simulations to evaluate conversational agents~\cite{svikhnushina-pu-2023-approximating, Afzali2023UserSimCRS}.
We expect our taxonomy to support a more realistic simulation of information-seeking conversations with conversational search engines.

%%%% The taxonomy suggest that users sometimes expect the engine to take initiative??

\subsection{Potential Considerations of Automatic Evaluation}

We chose to build an LLM-powered classifier to evaluate conversational search logs in an efficient and scalable way. 
We observed that our classifier yielded a moderate level of classification accuracy with enough coverage (Axis 1: 74.39\%, Axis 2: 71.11\%).

Still, we acknowledge that LLM-powered evaluation could lead to several potential problems.  
First, LLMs could lead to biased evaluation. 
For example, Ashwin et al.~\cite{ashwin2023using} showed that using LLMs for annotating interview transcripts could lead to biased annotations. They pointed out that the usage of LLM classification may not be suited for nuanced and context-specific tasks. 
To prevent potential biases in our context of search evaluation, future work could investigate classification and evaluation with LLMs incorporating contextual information. For example, considering users' prior search histories or query patterns could mitigate the effects of the innate characteristics of each user in the taxonomy classification results.  

Second, the stochastic nature of LLMs may make it challenging to produce consistent results.
In the classifier iteration process, we ran the classifier multiple times and used the majority voting to make results more consistent, but it may require a significant amount of time and resources. 
Furthermore, the black-box nature of LLMs could cause problems with diagnosing the causes of the misclassification. 
During our classifier iteration, we used the LLM to explain its reasons for misclassification. 
Future advances in the explainability of the LLMs could support the users of the classifier to determine whether classification results would be consistent to be trusted and to diagnose the misclassifications in a more reliable way. 

Lastly, as we used a fixed set of taxonomy, the classifier could fail to capture users' evolving usage of search engines. It would be necessary for the users to monitor the classifier results and make necessary updates consistently. Furthermore, we expect that the future version of the classifier could also support identifying new conversational patterns, as in LLM-supported inductive coding~\cite{Gao2023collabcoder}.

\subsection{Limitation \& Future Work}
% Our work has several limitations. 
First, we only focused on the relationship between two consecutive queries, potentially overlooking the complex structures of whole conversations. For example, our analysis method could not capture how users refine their search intents across multiple turns.
% For example, there were many cases where the user compiles how the search intents develop over conversational turns and specifies them in their new query, but such behaviors were not possible to capture with our approach. 
Future work could develop methodologies to understand the full context of conversations better.

Second, we collected the satisfaction ratings from the external evaluators rather than the users themselves. While rating by external evaluators yielded more objective evaluation results, it might deviate from the thoughts and search intents of the actual users. We expect that further research could study how our taxonomy relates to the search users' evaluation of the search session and examine whether the correlation differs from the findings of the current study. 

Third, while we used diverse types of search tasks to capture realistic use cases of search engines within our taxonomy, it could still be insufficient to capture all conversational patterns with search engines. Expanding the work with more diverse and naturalistic search tasks may uncover more fine-grained themes of the follow-up queries, leading to more insights into how follow-up query taxonomy could signal user satisfaction and other significant conversational quality attributes.

Lastly, as we developed the LLM classifier with logs from in-lab evaluation, we observed some limitations of the classifier with real-world conversation logs. In future work, we plan to further iterate the prompt with real-world data and use the fine-tuned model to improve its performance.

% \begin{acknowledgments}

% \end{acknowledgments}

%%
%% Define the bibliography file to be used
\bibliography{sample-ceur}

\appendix

\clearpage

\section{Appendix}

In this section, we present the prompts used for our classifier. The dataset utilized in this study is primarily in Korean therefore, the few-shot examples are original in Korean. However, parts of the classifier prompts and examples have been translated into English for clarity in this documentation. The whole code including the Korean prompt can be found at \texttt{https://github.com/kixlab/searchgpt-classifier}.

\subsection{Axis 1 Classifier}

\lstset{
  basicstyle=\ttfamily\footnotesize,
  columns=fullflexible,
  frame=single,
  breaklines=true,
  % postbreak=\mbox{\textcolor{red}{$\hookrightarrow$}\space},
  postbreak=\mbox{$\hookrightarrow$\space},
  breakatwhitespace=true,
  captionpos=b,
}

\begin{lstlisting}
You are a classifier analyzing an excerpt of a conversational search log, consisting of the initial query Q1, the engine's response R, and the follow-up query Q2. Among the taxonomy of the purpose of Q2 presented below, choose one that matches the best and only tell me the name of the taxonomy as this is scripted.

1. Clarifying Queries - Q2 clarifies the exact search intent of Q1. This occurs when Q2 is repeating the same question as Q1 to get a more specific answer by repeating the same question in Q1 with exact words or by asking the question about the same exact thing. If Q1 and Q2 are different, ask if the rewording of Q2 changes the meaning from Q1 or is just clarifying.

2. Exploring Domain - Q2 asks a question that explores the same general topic as Q1. Use taxonomy's definition widely.

3. Understanding Response - Q2 seeks more information necessary to understand R or to expand the range of information from R. This occurs when Q2 specifically asks for more information about information mentioned in R.

4. Narrowing Down - Q2 seeks more details or specific information that narrows down the question asked in Q1. This only occurs when Q2 asks for more specific information about something mentioned in Q1.

5. Seeking Different Representations - Q2 seeks different modalities (e.g., images, videos) or formats (e.g., tables) to better present the information provided in R.

6. Verifying Information - Q2 asks for more information to verify the information by asking for evidence, explanations, or alternatives from R

7. Reacting to Response - Q2 expresses satisfaction/dissatisfaction with R or provides feedback on R.

8. Unclassified - Q2 is completely irrelevant to Q1 or R.

Follow the examples as a guide to classify the queries. Assume the examples are always correct.

1:
Q1: What is the most preferred type of dishwasher by newlyweds?
R: Regarding dishwasher types, the most preferred product by newlyweds is a 12-person dishwasher. This capacity is generally suitable for a family of four. However, a 6-person...
Q2: Tell me the product name of the dishwasher preferred by newlyweds.

Taxonomy: Clarifying Queries

1A:
Q1: How can I apply for an additional government-supported rental loan if I already own a house?
R: The method to apply for an additional government-supported rental loan using your existing house as collateral is as follows...
Q2: I am curious about the income requirements for the government-supported rental loan. Tell me.

Taxonomy: Clarifying Queries

1B:
Q1: Tell me more about the Russian coup on June 24, 2023.
Q2: On June 24, 2023, tell me in detail about the military uprising led by the 'Wagner Group' in Russia.

Taxonomy: Clarifying Queries

2:
Q1: Is there any medicine I can take for a headache caused by a hangover?
R: There is no medicine you can take for a headache caused by a hangover. A hangover is the process of alcohol being metabolized...
Q2: What should I do if I accidentally took medicine for a severe hangover?

Taxonomy: Exploring Domain

2A:
Q1: Is it okay to bring toilet paper as a housewarming gift for newlyweds?
R: As for bringing toilet paper as a housewarming gift for newlyweds, it can be a good choice. However...
Q2: Tell me about a housewarming gift under 5,000 won that is not too expensive.

Taxonomy: Exploring Domain

2B:
Q1: So I shouldn't take headache medicine for a hangover?
Q2: Recommend a pain reliever with ibuprofen that helps with hangovers.

Taxonomy: Exploring Domain

3:
Q1: How about whiskey?
R: A whiskey product recommended as a housewarming gift for newlyweds in the 50,000 won range is available. Drinking it 'on the rocks' is a method of enjoying whiskey warmly, which can also be easily enjoyed at a newlywed's home. The product is an 'Oreforse' brand whiskey, which is popular for its luxury design and quality. Here are some recommended products:
Q2: What is 'on the rocks'?

Taxonomy: Understanding Response

4:
Q1: Choose a housewarming gift for newlyweds.
R: Suitable housewarming gifts for newlyweds include practical gifts like toilet paper, wet wipes, and coffee machines...
Q2: Choose a housewarming gift for newlyweds in the 50,000 won range.

Taxonomy: Narrowing Down

4A:
Q1: Tell me the product name of the dishwasher most preferred by newlyweds.
R: The dishwasher product most preferred by newlyweds is the BESPOKE dishwasher. This product offers various capacities and types...
Q2: How much is the BESPOKE dishwasher?

Taxonomy: Narrowing Down

4B:
Q1: What is the preferred capacity for a dishwasher?
R: The preferred capacity for a dishwasher is as follows...
Q2: Tell me the price of an LG dishwasher.

Taxonomy: Narrowing Down

5:
Q1: Heating cost fundraising poster
R: I cannot provide specific information about a heating cost fundraising poster, but generally, such posters include the following elements...
Q2: Show me a picture related to heating cost fundraising.

Taxonomy: Seeking Different Representations

6:
Q1: How do I remove polish from a steel tumbler, not stainless steel?
R: The method to remove polish from a steel tumbler is as follows...
Q2: Do I need to follow the steps from 1 to 6 in order?

Taxonomy: Verifying Information

6A:
Q1: Is the Wagner Group the largest army in Russia?
R: The largest army in Russia is the Russian Ground Forces, which is the largest in scale among the Russian military, with approximately 370,000 personnel. The Russian Ground Forces focus on mechanized and airborne units...
Q2: Isn't the Russian Ground Forces affiliated with the Russian government?

Taxonomy: Verifying Information

6B:
Q1: Can Tylenol help with a hangover?
R: Tylenol can help relieve headaches caused by hangovers. However, taking Tylenol after drinking can lead to liver damage...
Q2: Can I take a painkiller that contains acetaminophen to relieve a hangover?

Taxonomy: Verifying Information

7:
Q1: None of the three models you mentioned are in the 500,000 won range.
R: If I understand correctly, you want to know which of the three models I mentioned are not in the 500,000 won range...
Q2: I was talking about dishwashers, not Galaxy Tabs... Get it together, (LLM Agent Model name).

Taxonomy: Reacting to Response


\end{lstlisting}

\subsection{Axis 2 Classifier}
\begin{lstlisting}
You are a classifier analyzing an excerpt of a conversational search log, consisting of the initial query Q1, the engine's response R, and the follow-up query Q2. Among the taxonomy of the purpose of Q2 presented below, choose one Taxonomy within A, B, C, or D. Just list the name of the Theme. This taxonomy is based on priority. Check chronologically, so check if the conversation fits 1 then 2, then 3 and so on... The lower number gets priority.

A. Query-Specific (When Q2 is related to Q1):
1. Excluding Conditions - When the user explicitly removes specific conditions from the previous query (Q1).
2. Adding/Specifying Condition - WWhen the user specifies additional conditions or parameters to include in the query
3. Substituting Condition - When the user expresses the same query intent using different conditions or forms.

ONLY if Q2 is not related to Q1, then:
B. Response-Specific (When Q2 is unrelated to Q1 but related to R):
4. Converting Format - When Q2 asks in a different format (e.g., table, map, graph, image).
5. Confirming Response - When the user seeks clarification or confirms information from R using Q2.
6. Criticizing Response - When the user expresses criticism of the provided response (R).
7. Affirming Response - When the user expresses satisfaction with the provided response (R).
8. Requesting Additional Information - When the user requests additional information related to the response (R).

ONLY if Q2 is not related to Q1 and R, then:
C. Session-Specific (When Q2 is unrelated to Q1 and R but broadly to the topic):
9. Requesting Opinion - When the user asks about something an opinion not previously referred to in response (R).
10. Requesting Related Information - When the user requests related information that is broadly related to the same topic.

ONLY if Q2 is not related to Q1, R, and the topic, then:
D. Miscellaneous (Other):
11. Chatting Casually - When the user attempts chitchat rather than seeking specific information.
12. Requesting Unrelated Information.

Follow the examples as a guide to classify the queries. Assume the examples are always correct.

1:
Q1: What dishes can be made with beef, enoki mushrooms, shishito peppers, onions, and bean sprouts?
Q2: What dishes can be made with beef, enoki mushrooms, and onions?

Taxonomy: Excluding Conditions

2:
Q1: What is the most luxurious coffee machine?
Q2: Tell me about a more expensive and better coffee machine, excluding capsule coffee machines.

Taxonomy: Adding/Specifying Condition

2A:
Q1: Are there any models under 500,000 KRW from large corporations?
Q2: Not robot vacuums, but dishwashers!

Taxonomy: Adding/Specifying Condition

2B:
Q1: What kinds of special housewarming gifts are there for newlyweds?
Q2: Recommend types of foreign electronic products.

Taxonomy: Adding/Specifying Condition

2C:
Q1: Tell me about the Russian coup.
Q2: Tell me more about the recent coup related to the Wagner Group.

Taxonomy: Adding/Specifying Condition

3:
Q1: Recommend a restaurant in Daejeon suitable for a 70th birthday party with a rating of at least 4.8.
Q2: Recommend a well-reviewed restaurant in Daejeon suitable for a 70th birthday party.

Taxonomy: Substituting Condition

3A:
Q1: Tell me more about the second coup.
Q2: Tell me more about the Russian coup on June 24, 2023.

Taxonomy: Substituting Condition

5:
Q1: Can I bring toilet paper as a housewarming gift for newlyweds?
Q2: Tell me about a housewarming gift under 5,000 won that isn't too expensive.

Taxonomy: Criticizing Response

7:
Q1: Is the Wagner Group the largest army in Russia?
R: The largest army in Russia is the Russian Ground Forces, which is the largest among the Russian military, with approximately 370,000 personnel...
Q2: Isn't the Russian Ground Forces affiliated with the Russian government?

Taxonomy: Confirming Response

7A:
1: Can Tylenol help with a hangover?
R: Can Tylenol help with a hangover? - A hangover involves symptoms like headache, vomiting, and dizziness that occur after drinking alcohol...
Q2: So, can I take a pain reliever containing acetaminophen for hangover relief?

Taxonomy: Confirming Response

7B:
Q1: Russia doesn't have a military organization other than the Wagner Group, does it? 
Q2: Is the Wagner Group the largest army in Russia?

Taxonomy: Confirming Response

8:
Q1: Choose a housewarming gift for newlyweds in the 50,000 won range. 
R: Here are some recommended housewarming gifts for newlyweds. 1. Handmade pottery gifts for newlyweds. Q2: How about whiskey?

Taxonomy: Requesting Opinion

8A:
Q1: Is it okay to exercise when I have a severe hangover? 
R: Opinions vary on whether you can exercise with a severe hangover. However, some argue that you can still engage in some physical activity even with a severe hangover. 
Q2: Can I go to a bathhouse if I have a headache from a severe hangover?

Taxonomy: Requesting Opinion

9:
Q1: I'm planning a ryokan trip, any good places?
R: Are there any good ryokan places for a trip? - Many ryokans in the Yufuin area are frequently searched. Some ryokans are praised for their breakfast,...
Q2: Are there other places worth visiting in Yufuin?

Taxonomy: Requesting Additional Information

\end{lstlisting}

\end{document}